\documentclass[showpacs,preprintnumbers,showkeys,amsmath,amssymb]{revtex4}
\usepackage{graphicx}
\usepackage{dcolumn}
\usepackage{bm}
\begin{document}

\preprint{}

\title{Implementations of two-photon four-qubit Toffoli and Fredkin gates assisted by nitrogen-vacancy centers}

\author{Hai-Rui Wei\footnote{Corresponding author:hrwei@ustb.edu.cn} and Pei-Jin Zhu }

\address{School of Mathematics and Physics, University of Science and Technology Beijing, Beijing 100083, China}

\begin{abstract}
It is desirable to implement an efficient quantum information process demanding fewer quantum resources.  We designed two compact quantum circuits for determinately implementing four-qubit Toffoli and Fredkin gates on single-photon systems in both the polarization and spatial degrees of freedom (DoFs) via diamond nitrogen-vacancy (NV) centers in resonators. The gates are heralded by the electron spins associated with the diamond NV centers. In contrast to the ones with one DoF, our implementations reduce the quantum resource and are robust against the decoherence. Evaluations of fidelities and efficiencies of our gates show that our schemes may be implemented with current technology.
\end{abstract}



\maketitle

\section{Introduction}

Quantum computer offers great advantages over the classical computer\cite{Book}, such as tremendous speedup, efficiently searching for unordered database, and factorizing large numbers. Quantum gates are the fundamental elements for the quantum computer. Quantum computation, quantum networks, and quantum algorithms can be decomposed into a set of universal gates and single-qubit gates\cite{network}. Controlled-NOT (CNOT) gate or controlled-phase gate is the most popular nontrivial universal gate\cite{Universal}, and considerable progresses have been made on them\cite{progress1,progress2,long1}. In recent years, constructions of the Toffoli gate\cite{Toffoli} or Fredkin gate\cite{Fredkin} has been addressed from various perspectives, and these two universal multiqubit-control gates are the key operations in quantum networks, quantum phase estimate, fault-error quantum computing, quantum-error correction, quantum Shor's algorithm\cite{Shor1,Shor2}, quantum Grover's algorithm\cite{Grover1}, and optimal Long's algorithm\cite{Grover2}. A Toffoli or Fredkin gate can be implemented by series of nontrivial two-qubit gates along with single-qubit gates. However, this strategy requires a number of gate steps, additional qubits, and increases the noise channels.  The synthesis of a $n$-qubit Toffoli gate requires $O(n^2)$ two-qubit gates\cite{Universal}. The minimal cost of a three-qubit Fredkin gate is five two-qubit gates\cite{Fredkin-cost}.  Therefore, it is significant to directly and efficiently implemented  Toffoli and Fredkin gates without resorting to two- and single-qubit gates.

Finding an efficient quantum computing with minimal quantum resources is the central work in quantum information process. Encoding  qubits in multiple degrees of freedom (DoFs) increases the capacity of the quantum channel, reduces the quantum resources, and are robust against the decoherence caused by the environments. Single photon is nowadays recognized as an excellent candidate for quantum information process in multiple DoFs as it carries many qubit-like DoFs\cite{multiple} including momentum (linear, orbital, spatial), frequency, time-bin, polarization, and availability single-qubit manipulation. Utilizing cross-Kerr nonlinearity, Sheng \emph{et al.}\cite{hyper-Bell-Sheng} and Liu \emph{et al.}\cite{hyper-Bell-Liuqian} proposed schemes for complete hyperentangled-Bell-state analysis in two and three DoFs, respectively. In  2012, Ren \emph{et al.}\cite{hyper-Bell-Ren} theoretically implemented hyperentangled-Bell-state analysis in polarization-spatial DoF via quantum dot. In 2014, Wang \emph{et al.}\cite{hyperdistillation} proposed a hyperdistillation scheme in two DoFs. In 2015, Wang \emph{et al}.\cite{pan} demonstrated quantum teleportation of a single photon with polarization-orbit DoF. Recently, Scholz \emph{et al}.\cite{Scholz-DJ} and Zhang \emph{et al}.\cite{Zhang-DJ} demonstrated two-qubit Deutsch-Jozsa algorithm with a single photon in polarization-spatial DoF, respectively. Abouraddy \emph{et al}.\cite{one-photon-three-qubit} constructed one-photon three-qubit universal gates. The weak interaction in single photon level is the limitation of photonic quantum computing. Fortunately, in 2001, Knill, Laflamme, and Milburn\cite{KLM} proposed an interesting scheme for implementing linear optical controlled-phase gate with maximal success probability of 3/4. The strategy adopted to overcome undeterministic is to employ cross-Kerr nonlinearity\cite{kerr1} or photon-matter platforms\cite{platform1,platform2}. Up to now, the giant cross-Kerr nonlinearity is still a challenge in experiment. In recent years, wide attention has been drawn to couple flying photonic qubit to stationary matter qubit as it provides a platform for scalable photonic or matter quantum computing. Ren \emph{et al.}\cite{hyper-CNOT1,hyper-CNOT2} proposed a spatial-polarization hyper-controlled-not gate with coupled quantum dot. Ren \emph{et al.}\cite{Hyper-Toffoli-Ren}, Wei \emph{et al.}\cite{Hyper-Toffoli-Wei}, and Wang \emph{et al.}\cite{Hyper-Toffoli-Wang} presented schemes for multiphoton hyper-parallel quantum computing assisted by matter qubits coupled with resonators.

Isolated electron-spin in diamond nitrogen-vacancy (NV) center has been recognized as an unique and promising candidate for solid-state quantum computing. The appeal feature of the diamond NV center is its urtralong coherence time ($\sim m$s) even at room temperature\cite{coherence1,coherence2}, and diamond NV center has been received strong interest in recent years. The initialization\cite{population}, manipulation\cite{manipulation}, storage\cite{storage}, readout\cite{readout}, and the nontrivial interactions between two diamond NV spins\cite{Hanson1,Hanson2}, which are the necessary operations for quantum computing, have been well experimentally demonstrated. There are also a number of hallmark other demonstrations, such as the quantum Deutsch-Jozsa algorithm\cite{algorithm}, controlled-ROT gate\cite{controlled-ROT}, decoherence-protected controlled-rotation gates\cite{protected-gate}, and holonomic CNOT gate\cite{geometric1,geometric2}. Based on photon-NV entangled platform\cite{photon-NV-PRL1,photon-NV-PRL2}, Hanson \emph{et al}.\cite{Hanson2,Hanson3} experimentally demonstrated quantum teleportation and Bell inequality between distant diamond NV centers, and several proposals for implementing parallel and hyper-parallel quantum computing were also proposed\cite{Hyper-Toffoli-Wang,ourphotongate-NV,ourgates-NV,Wangchuan-NV}.

In this article, we investigate the implementations of the four-qubit Toffoli and Fredkin gates acting on two-photon systems via the diamond NV centers coupled to resonators. The Toffoli gate flips the polarization or spatial state of the target qubit conditional on the polarization and spatial states of the control qubits. The Fredkin gate swaps the polarization and spatial states of the target photon depending on the polarization and spatial states of the control photon. In our schemes, the two gates are heralded by the spins of the electrons in diamond NV centers. Additional  photons, necessary for the cross-Kerr-based and parity-check-based optical quantum computing, are not required. Compared to the ones in one DoF, our schemes only require the half quantum resource, and are robust against the decoherence. The estimations of our devices show that our schemes are possible in experiment with current state of technology.

\section{Results}\label{sec2}
{\bf The optical property of an NV-cavity platform.}
 The Toffoli gates we considered are equivalent to the controlled-controlled-controlled-phase flip (CCCPF) gate up to two Hadamard  ($H$) gates acting on the target qubits, i.e., $U_{\text{Toffoli}}=H.\;\text{CCCPF}.\;H$. Here, we employ the NV-cavity entangled platform to overcome the weak interactions between two single photons involved in the Toffoli gate. As shown in Fig. 1(a),  $|m_s\rangle=|\pm 1\rangle$ and $|m_s\rangle=|0\rangle$ (marked by $|\pm\rangle,\;|0\rangle$) are the ground triple states of the diamond NV center. In absence of a magnetic field, $|\pm 1\rangle $ and $|0\rangle$ are split with 2.87 GHz due to the spin-spin interaction, and the corresponding transitions, $|0\rangle\leftrightarrow|\pm1\rangle$, are in the microwave regime. The six electron excited states include\cite{excited-state}
$|E_1\rangle=|E_-\rangle|-1\rangle-|E_+\rangle|+1\rangle$,
$|E_2\rangle=|E_-\rangle|-1\rangle+|E_+\rangle|+1\rangle$,
$|E_x\rangle=|X\rangle|0\rangle$,
$|E_y\rangle=|Y\rangle|0\rangle$,
$|A_1\rangle =|E_-\rangle|+1\rangle-|E_+\rangle|-1\rangle$, and
$|A_2\rangle=|E_-\rangle|+1\rangle+|E_+\rangle|-1\rangle$. $|E_{\pm}\rangle$, and $|X\rangle$, $|Y\rangle$ are orbital states with angular momentum projections $\pm1$ and $0$, respectively.
The energy gap arising from the spin-spin and spin-orbit interactions protects state $|A_2\rangle$ against small strain and magnetic filed. The transitions, $|\pm\rangle\rightarrow|A_2\rangle$, are driven by the resonator optical pluses with $\sigma_-$ and $\sigma_+$-polarized at 637 nm, respectively (see Fig. 1(b)). Therefore, the device shown in Fig. 1 can be used as a quantum interface between optical and solid-state systems for scalable quantum information process.

\begin{figure}[!h]       
\centering
\includegraphics[width=8.5 cm,angle=0]{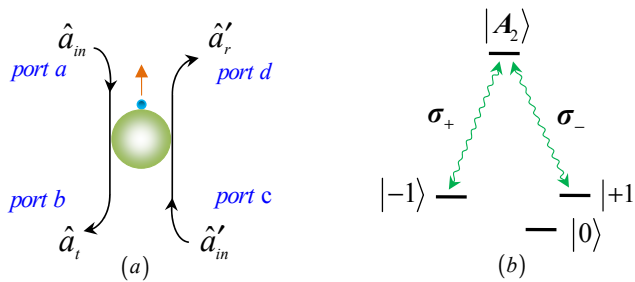}
\caption{(a) Schematic diagram for a diamond NV center coupled to a double-sided resonator. (b) A $\wedge$-type electron energy structure of a trapped diamond NV center. Transitions $|\pm1\rangle\leftrightarrow|A_2\rangle$ are driven by the $\sigma_-$ and $\sigma_+$  photons, respectively.
}
\label{Level}
\end{figure}

Compared with the coupling of an unconfined diamond NV center, trapping a diamond NV center into an optical cavity enhances the coupling between the incident photon and the diamond NV center. Nowadays, various of optical microcavities, such as photonic crystal cavity\cite{crystal}, fiber-based Fabry-Perot cavity\cite{fiber-based}, microsphere\cite{microsphere}, toroidal resonators\cite{toroidal}, coupling to the diamond NV centers have been experimentally demonstrated. Suppose that the diamond NV center coupling with a whispering-gallery-mode microresonator with two tapered fibers (see Fig. 1). If the diamond NV in the state $|+1\rangle$, the incoming $\sigma_+$-polarized photon via port $a$ (port $c$) senses a cold cavity, and departs through port $b$ ($d$) with phase shift $e^{i\varphi_0}$, while the incoming $\sigma_-$-polarized photon via port $a$ (port $c$) senses a hot cavity, and leaves from the cavity via port $d$ ($b$) with phase shift $e^{i\varphi_h}$. In the same way, If the diamond NV center in the state $|-\rangle$, the input $\sigma_-$-polarized photon senses a cold cavity and it is transmitted, while the $\sigma_+$-polarized photon senses the hot cavity and it is reflected. The reflection and transmission coefficients of the NV-cavity system can be obtained by solving the Heisenberg equations for the cavity mode $a$ with frequency $\omega_{c}$ and the diamond NV lowering operator $\sigma_{-}$ with frequency $\omega_{0}$, that is\cite{Heisenberg},
\begin{eqnarray}       \label{eq1}
\begin{split}
&\dot{a} = -\left[i(\omega_c-\omega_p)+\kappa+\frac{\kappa_s}{2}\right]a-g\sigma_{-} - \sqrt{\kappa}(a_{\rm{in}}+a'_{\rm{in}})-g\sigma_{-}-h, \\
&\dot{\sigma}_{-} =-\left[i(\omega_{0}-\omega_p)+\frac{\gamma}{2}\right]\sigma_{-}-g\sigma_z a- f.
\end{split}
\end{eqnarray}
Here, $\kappa$ and $\gamma/2$ are the decay rates of the cavity field and the diamond NV dipole, respectively. $\kappa_s/2$ is the cavity intrinsic loss rate (side leakage). $g$ is the coupling rate between the diamond NV center and the resonator. $\sigma_z$ is the pauli operator. $h$ and $f$ are the noise operators. The input field is connected with the output filed by $a_{\rm{r}}=a_{\rm{in}}+\sqrt{\kappa}a$ and $a_t=a_{\rm{in}}'+\sqrt{\kappa}a$. Taking the weak excitation limitation, i.e., $<\sigma_z>\approx-1$, through the operation, the transmission and reflection coefficients of the NV-cavity system can be respectively written as\cite{An,Hu2009}
\begin{eqnarray}       \label{eq2}
\begin{split}
&t(\omega_p) = - \frac{-\kappa\big[i(\omega_0-\omega_p)+\frac{\gamma}{2}\big]}{\big[i(\omega_0-\omega_p)+\frac{\gamma}{2}\big]\big[i(\omega_c-\omega_p)+\kappa+\frac{\kappa_s}{2}\big]+g^2},\\
&r(\omega_p) =\frac{\big[i(\omega_0-\omega_p)+\frac{\gamma}{2}\big]\big[i(\omega_c-\omega_p)+\frac{\kappa_s}{2}\big]+g^2}{\big[i(\omega_0-\omega_p)+\frac{\gamma}{2}\big]\big[i(\omega_c-\omega_p)+\kappa+\frac{\kappa_s}{2}\big]+g^2}.
\end{split}
\end{eqnarray}
Under the resonator condition $\omega_c=\omega_0=\omega_p$, if the Purcell factor $g^2/(\kappa\gamma)\gg1$ with $\kappa_s\approx0$ is ignored, the cavity mode of the hot cavity is reflected, i.e., $r(\omega_p)\rightarrow1$ and $t(\omega_p)\rightarrow0$. If $\kappa_s/\kappa\ll1$, the cavity mode of the bare cavity is transmission, i.e., $r_0(\omega_p)\rightarrow0$ and $t_0(\omega_p)\rightarrow-1$. That is, the input-output relations between the single photon and NV center can be summarized as
\begin{eqnarray}       \label{eq3}
\begin{split}
&|R,a\rangle|+1\rangle\;\rightarrow\;|L,d\rangle|+1\rangle,\quad\qquad|R,a\rangle|-1\rangle\;\rightarrow\;-|R,b\rangle|-1\rangle,\\
&|L,c\rangle|+1\rangle\;\rightarrow\;|R,b\rangle|+1\rangle,\quad\qquad|L,c\rangle|-1\rangle\;\rightarrow\;-|L,d\rangle|-1\rangle,\\
&|R,c\rangle|+1\rangle\;\rightarrow\;-|R,d\rangle|+1\rangle,\;\qquad|R,c\rangle|-1\rangle\;\rightarrow\;|L,b\rangle|-1\rangle,\\
&|L,a\rangle|+1\rangle\;\rightarrow\;-|L,b\rangle|+1\rangle,\;\qquad|L,a\rangle|-1\rangle\;\rightarrow\;|R,d\rangle|-1\rangle.
\end{split}
\end{eqnarray}
Here, $|R,a\rangle$ and $|L,d\rangle$ with $s_z=-1$, $|L,a\rangle$ and $|R,d\rangle$ with $s_z=+1$, the polarization of the photon is defined with respect to the $z$ axis (the diamond NV center axis).


{\bf Two-photon four-qubit Toffoli gate acting on the polarization and spatial DoFs.}
%
%

The quantum circuit we designed, as shown in Fig. 2, performs the following operation on a two-photon four-qubit state
\begin{eqnarray}       \label{eq4}
\begin{split}
|\psi\rangle=&\;(\alpha_{1}|R_1\rangle + \alpha_{2}|L_1\rangle) \otimes (\beta_{1}|a_1\rangle + \beta_{2}|a_2\rangle)
\otimes (\gamma_{1}|R_2\rangle + \gamma_{2}|L_2\rangle) \otimes (\delta_{1}|b_1\rangle + \delta_{2}|b_2\rangle)
\\ \rightarrow&\;
[\alpha_{1} \beta_{1}|R_1\rangle|a_1\rangle
+\alpha_{1} \beta_{2}|R_1\rangle |a_2\rangle
+\alpha_{2}\beta_{1}|L_1\rangle|a_1\rangle](\gamma_{1}|R_2\rangle +\gamma_{2}|L_2\rangle)(\delta_{1}|b_1\rangle + \delta_{2}|b_2\rangle)
  \\&
+\alpha_{2}\beta_{2}\gamma_{1}|L_1\rangle |a_2\rangle|R_2\rangle (\delta_{1}|b_1\rangle + \delta_{2}|b_2\rangle)+ \alpha_{2}\beta_{2}\gamma_{2}|L_1\rangle|a_2\rangle|L_2\rangle (\delta_{1}|b_1\rangle - \delta_{2}|b_2\rangle),
\end{split}
\end{eqnarray}
where $\alpha_1$, $\alpha_2$, $\beta_1$, $\beta_2$, $\gamma_1$, $\gamma_2$, $\delta_1$, and $\delta_2$ are arbitrary complex parameters satisfying  $|\alpha_1|^2+|\alpha_2|^2=1$, $|\beta_1|^2+|\beta_2|^2=1$, $|\gamma_1|^2+|\gamma_2|^2=1$, and $|\delta_1|^2+|\delta_2|^2=1$. The indices 1 and 2 of $R$ (or $L$) denote the $R$- (or $L$-) polarized wave-packets of the photon 1 and photon 2, respectively.  $a_1$ ($b_1$) and $a_2$ ($b_2$) represent photon 1 (photon 2) emitted from the spatial modes 1 and 2, respectively.

\begin{figure}
\centering
\includegraphics[width=6.5 cm,angle=0]{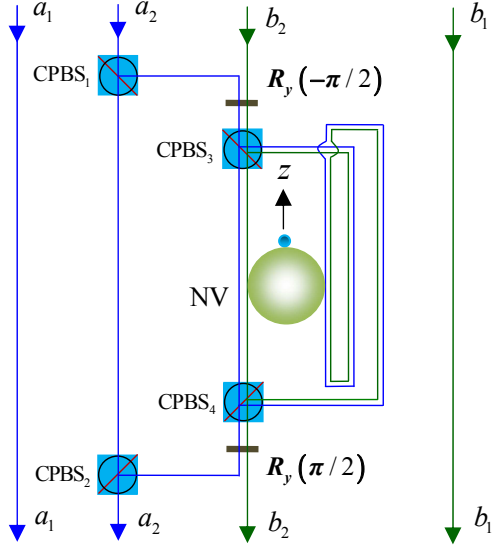}
\caption{Two-photon four-qubit Toffoli gate acting on the polarization and spatial DoFs. Polarizing beam splitters in the circularly basis, CPBS$_i$ ($i=1,2,3,4$),  transmits
the $R$-polarized and reflects the $L$-polarized components, respectively. Single-qubit operations $R_y(\pm\pi/2)=e^{\mp i\frac{\pi}{4}\sigma_y}$ can be achieved by employing two half-wave plates oriented at $45^\circ$ and $22.5^\circ$, respectively. The local single-qubit operations, heralded by the outcomes the diamond NV center, via a feed-forward process are not shown in the sketch.
}
 \label{Toffoli}
\end{figure}

In Fig. 2, we first use the circularly polarized beam splitter, CPBS$_1$, to depart the wave-packets, $R$ and $L$, of photon 1 emitted from the spatial mode $a_2$, i.e., the $R_1$-component is transmitted and the $L_1$-component is reflected. The $R_1$-component does not interact with the diamond NV center and arrives at CPBS$_2$ directly. Before and after the $L_1$-component interacts with the block composed of CPBS$_3$, NV, and CPBS$_4$, the two single-qubit rotations $R_y(-\pi/2)$ and $R_y(\pi/2)$ are respectively applied on photon 1 simultaneously with two Hadamard operations $H_e$s applied on the diamond NV center. Subsequently, the $R_1$-component and $L_1$-component mix at CPBS$_2$. The operations ($\text{CPBS}_1 \rightarrow R_y(-\pi/2), H_e \rightarrow \text{CPBS}_3 \rightarrow \text{NV} \rightarrow \text{CPBS}_4 \rightarrow R_y(\pi/2), H_e \rightarrow \text{CPBS}_2$) transform the system composed of the photon 1, photon 2, and diamond NV center from the initial state $|\psi_0\rangle$ into $|\psi_1\rangle$. Here,
\begin{eqnarray}       \label{eq5}
|\psi_0\rangle=(\alpha_{1}|R_1\rangle + \alpha_{2}|L_1\rangle) \otimes (\beta_{1}|a_1\rangle + \beta_{2}|a_2\rangle)
\otimes (\gamma_{1}|R_2\rangle + \gamma_{2}|L_2\rangle) \otimes (\delta_{1}|b_1\rangle + \delta_{2}|b_2\rangle) \otimes |-\rangle,
\end{eqnarray}
\begin{eqnarray}       \label{eq6}
\begin{split}
|\psi_1\rangle=\;&
(\alpha_{1}\beta_{1}|R_1\rangle|a_1\rangle|-\rangle
+\alpha_{1}\beta_{2}|R_1\rangle|a_2\rangle|-\rangle
+\alpha_{2}\beta_{1}|L_1\rangle|a_1\rangle|-\rangle
+\alpha_{2}\beta_{2}|L_1\rangle|a_2\rangle|+\rangle)
 (\gamma_{1}|R_2\rangle + \gamma_{2}|L_2\rangle) \qquad\;\\&\otimes (\delta_{1}|b_1\rangle + \delta_{2}|b_2\rangle) .
\end{split}
\end{eqnarray}
The Hadamard transformation $H_e$ can be achieved by employing $\pi/2$ pulse and it completes the following transformations
\begin{eqnarray}       \label{eq7}
|+\rangle \xrightarrow{H_e} \frac{1}{\sqrt{2}}(|+\rangle+|-\rangle),  \qquad
|-\rangle \xrightarrow{H_e} \frac{1}{\sqrt{2}}(|+\rangle-|-\rangle).
\end{eqnarray}
The single-qubit rotations $R_y(\pm\pi/2)$ can be achieved by employing two half-wave plates oriented at $45^\circ$ and $22.5^\circ$, respectively. In the standard basis $\{|R\rangle,\;|L\rangle\}$,  $R_y(\pm\pi/2)$  can be written as
\begin{eqnarray}       \label{eq8}
R_y(+\frac{\pi}{2})=\frac{1}{\sqrt{2}}
\left( \begin{array}{cc}
1 & -1 \\
1 & 1 \\
 \end{array}
 \right),  \quad
 R_y(-\frac{\pi}{2})=\frac{1}{\sqrt{2}}
\left( \begin{array}{cc}
1 & 1 \\
-1 & 1 \\
 \end{array}
 \right).
\end{eqnarray}

Next, photon 2 in the spatial mode $b_2$ is injected into the block. Before and after photon 2 interacts with the block, $R_y(\mp\pi/2)$  are applied on it, respectively. The above operations ($ R_y(-\pi/2) \rightarrow \text{CPBS}_3 \rightarrow \text{NV} \rightarrow \text{CPBS}_4 \rightarrow R_y(\pi/2)$) transform $|\psi_1\rangle$ into
\begin{eqnarray}       \label{eq9}
\begin{split}
|\psi_2\rangle=\;&
\alpha_{1}|R_1\rangle(\beta_{1}|a_1\rangle+\beta_{2}|a_2\rangle)(\gamma_{1}|R_2\rangle + \gamma_{2}|L_2\rangle)(\delta_{1}|b_1\rangle - \delta_{2}|b_2\rangle)|-\rangle\\&
+\alpha_{2}\beta_{1}|L_1\rangle|a_1\rangle(\gamma_{1}|R_2\rangle + \gamma_{2}|L_2\rangle)(\delta_{1}|b_1\rangle - \delta_{2}|b_2\rangle)|-\rangle\\&
+\alpha_{2}\beta_{2}|L_1\rangle|a_2\rangle\big[\gamma_{1}|R_2\rangle (\delta_{1}|b_1\rangle - \delta_{2}|b_2\rangle)
                                               + \gamma_{2}|L_2\rangle (\delta_{1}|b_1\rangle + \delta_{2}|b_2\rangle)\big]|+\rangle.
\end{split}
\end{eqnarray}

Third, we measure the electron spin state of the diamond NV center in the basis  $\{(|+\rangle\pm|-\rangle)/\sqrt{2}\}$. If the diamond NV center in the state  $(|+\rangle+|-\rangle)/\sqrt{2}$, we perform a phase shift $\pi$ on the photon emitted from the spatial mode $b_2$, i.e., $|R_2\rangle|b_2\rangle \rightarrow -|R_2\rangle|b_2\rangle $ and $|L_2\rangle|b_2\rangle \rightarrow -|L_2\rangle|b_2\rangle$. If the diamond NV  in the state  $(|+\rangle-|-\rangle)/\sqrt{2}$,  classical feed-forward operations, phase shift $\pi$ and  $\sigma_z=|R\rangle\langle R|-|L\rangle\langle L|$, are performed on the photons emitted from the spatial modes $b_2$ and $a_2$, respectively.  After the above operations, the state of the system composed the two photons becomes
\begin{eqnarray}       \label{eq10}
\begin{split}
|\psi_3\rangle=\;&
(\alpha_{1}\beta_{1}|R_1\rangle|a_1\rangle+\alpha_{1}\beta_{2}|R_1\rangle|a_2\rangle+\alpha_{2}\beta_{1}|L_1\rangle|a_1\rangle)(\gamma_{1}|R_2\rangle + \gamma_{2}|L_2\rangle)(\delta_{1}|b_1\rangle + \delta_{2}|b_2\rangle)\\&
+\alpha_{2}\beta_{2}|L_1\rangle|a_2\rangle\big[\gamma_{1}|R_2\rangle (\delta_{1}|b_1\rangle + \delta_{2}|b_2\rangle)
                                               + \gamma_{2}|L_2\rangle (\delta_{1}|b_1\rangle - \delta_{2}|b_2\rangle)\big].
\end{split}
\end{eqnarray}

It is obvious that the quantum circuit shown in Fig. 2 determinately implementing a CCCPF gate which flips the phase of the input state if and only if the system initially is in the state $|L_1\rangle|a_2\rangle|L_2\rangle|b_2\rangle$, and has no change otherwise. Therefore, a two-photon four-qubit Toffoli gate which performs a NOT operation on the spatial (or polarization) state of the photon 2 depending on the states of other qubits can be achieved by the scheme depicted by Fig. 2 up to two Hadamard gates.


{\bf Two-photon four-qubit Fredkin gate acting on the polarization and spatial DoFs.}
The Fredkin gate swaps the states of the two target qubits conditional on the states of the control qubits. The quantum circuit depicted by Fig. 3 implements a Fredkin gate which completes the transformation
\begin{eqnarray}       \label{eq11}
\begin{split}
|\varphi\rangle=&\;(\alpha_{1}|R_1\rangle + \alpha_{2}|L_1\rangle) \otimes (\beta_{1}|a_1\rangle + \beta_{2}|a_2\rangle)
\otimes (\gamma_{1}|R_2\rangle + \gamma_{2}|L_2\rangle) \otimes (\delta_{1}|b_1\rangle + \delta_{2}|b_2\rangle)
  \\ \rightarrow&\;
\alpha_{1}|R_1\rangle (\beta_{1}|a_1\rangle + \beta_{2} |a_2\rangle)(\gamma_{1}|R_2\rangle + \gamma_{2}|L_2\rangle)(\delta_{1}|b_1\rangle + \delta_{2}|b_2\rangle)
\\ &+\alpha_{2}\beta_{1} |L_1\rangle |a_1\rangle (\gamma_{1}|R_2\rangle + \gamma_{2}|L_2\rangle)(\delta_{1}|b_1\rangle + \delta_{2}|b_2\rangle)
+\alpha_{2}\beta_{2}|L_1\rangle|a_2\rangle(\delta_{1}|R_2\rangle + \delta_{2}|L_2\rangle)(\gamma_{1}|b_1\rangle + \gamma_{2}|b_2\rangle).
\end{split}
\end{eqnarray}
As illustrated in Fig. 3, such Fredkin gate can be implemented by three steps. We first inject the photon 1 (encoded as the control photon) emitted from the spatial mode $a_2$ into the block composed of CPBS$_3$, phase shifters $P_{\pi_1}$  and $P_{\pi_2}$, NV, and CPBS$_4$. Before and after the photon interacts with such block, Hadamard operations, $H_{p_1}$ and $H_{p_2}$, are respectively applied on it with the half-wave plates oriented at 22.5$^\circ$ simultaneously with two Hadamard operations $H_e$s  applied on the diamond NV centers. The above operations ($\text{CPBS}_1 \rightarrow H_{p_1}, H_e \rightarrow \text{CPBS}_3 \rightarrow P_{\pi_1} \rightarrow \text{NV} \rightarrow P_{\pi_2} \rightarrow \text{CPBS}_4 \rightarrow H_{p_2}, H_e \rightarrow \text{CPBS}_2$) transform the system composed of the two single photons and one diamond NV center from the initial state $|\varphi_0\rangle$ into $|\varphi_1\rangle$. Here,
\begin{eqnarray}       \label{eq12}
|\varphi_0\rangle=(\alpha_{1}|R_1\rangle + \alpha_{2}|L_1\rangle) \otimes (\beta_{1}|a_1\rangle + \beta_{2}|a_2\rangle)
\otimes (\gamma_{1}|R_2\rangle + \gamma_{2}|L_2\rangle) \otimes (\delta_{1}|b_1\rangle + \delta_{2}|b_2\rangle) \otimes |-\rangle,
\end{eqnarray}
\begin{eqnarray}       \label{eq13}
\begin{split}
|\varphi_1\rangle=\;&
\big[\alpha_{1}|R_1\rangle (\beta_{1}|a_1\rangle|-\rangle + \beta_{2}|a_2\rangle|-\rangle)
+\alpha_{2}|L_1\rangle(\beta_{1}|a_1\rangle |-\rangle - \beta_{2}|a_2\rangle|+\rangle)\big]
(\gamma_{1}|R_2\rangle + \gamma_{2}|L_2\rangle) (\delta_{1}|b_1\rangle + \delta_{2}|b_2\rangle).
\end{split}
\end{eqnarray}
In the standard basis $\{|R\rangle,|L\rangle\}$, Hadamard operation $H_p$ can be expressed as
\begin{eqnarray}       \label{eq14}
H_p=\frac{1}{\sqrt{2}}
\left( \begin{array}{cc}
1 & 1 \\
1 & -1 \\
 \end{array}
 \right).
\end{eqnarray}
Phase shifter, $P_{\pi_1}$ or $P_{\pi_2}$, introduces a phase shift  $\pi$ to the passing photon, i.e., $|R\rangle\rightarrow-|R\rangle$ and $|L\rangle\rightarrow-|L\rangle$.

\begin{figure}
\centering
\includegraphics[width=7.5 cm,angle=0]{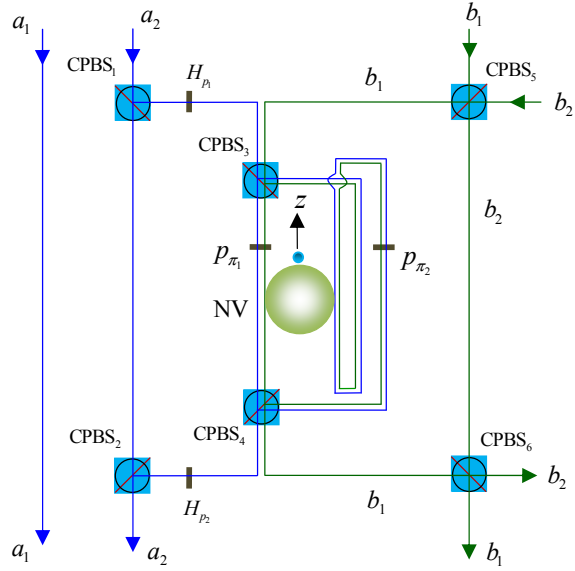}
\caption{Two-photon four-qubit Fredkin gate acting on the polarization and spatial DoFs. Hadamard transformation $H_j$ can be achieved by employing a half-wave plate oriented at $22.5^\circ$. Phase shifter $P_{\pi}$ completes the operations $|R\rangle\rightarrow-|R\rangle$ and $|L\rangle\rightarrow-|L\rangle$. The local classical feed-forward single-qubit operations are not shown in the sketch.
}
 \label{Fredkin}
\end{figure}

Subsequently, photon 2 emitted form the spatial mode $b_1$ or $b_2$ is injected and then arrives at CPBS$_5$. CPBS$_5$ transforms $|\varphi_1\rangle$ into
\begin{eqnarray}       \label{eq15}
\begin{split}
|\varphi_3\rangle=\;&
\big[\alpha_{1}|R_1\rangle (\beta_{1}|a_1\rangle + \beta_{2}|a_2\rangle)|-\rangle
+\alpha_{2}|L_1\rangle(\beta_{1}|a_1\rangle |-\rangle - \beta_{2}|a_2\rangle|+\rangle)\big] \\&\otimes
(\gamma_{1}\delta_{1}|R_2\rangle|b_2\rangle + \gamma_{1}\delta_{2}|R_2\rangle|b_1\rangle
 +\gamma_{2}\delta_{1}|L_2\rangle|b_1\rangle + \gamma_{2}\delta_{2}|L_2\rangle|b_2\rangle).
\end{split}
\end{eqnarray}
The wave-packets emitted from the spatial mode $b_1$ pass through the block composed of CPBS$_3$, phase shifters $P_{\pi_1}$  and $P_{\pi_2}$, NV, and CPBS$_4$, and such block transforms the state of the whole system into
\begin{eqnarray}       \label{eq16}
\begin{split}
|\varphi_4\rangle=\;&
(\alpha_{1}\beta_{1}|R_1\rangle|a_1\rangle+\alpha_{1}\beta_{2}|R_1\rangle|a_2\rangle+\alpha_{2}\beta_{1}|L_1\rangle|a_1\rangle)\\&\otimes
 (\gamma_{1}\delta_{1}|R_2\rangle|b_2\rangle+\gamma_{1}\delta_{2}|R_2\rangle|b_1\rangle
+ \gamma_{2}\delta_{1}|L_2\rangle|b_1\rangle+\gamma_{2}\delta_{2}|R_2\rangle|b_2\rangle)|-\rangle\\&
-\alpha_{2}\beta_{2}|L_1\rangle|a_2\rangle
(\gamma_{1}\delta_{1}|R_2\rangle|b_2\rangle
+\gamma_{1}\delta_{2}|L_2\rangle|b_1\rangle
+\gamma_{2}\delta_{1}|R_2\rangle|b_1\rangle
+\gamma_{2}\delta_{2}|L_2\rangle|b_2\rangle)|+\rangle.
\end{split}
\end{eqnarray}
After the wave-packets of  photon 2 mix at CPBS$_6$, the state of the whole system becomes
\begin{eqnarray}       \label{eq17}
\begin{split}
|\varphi_5\rangle=\;&
\alpha_{1}|R_1\rangle(\beta_{1}|a_1\rangle+\beta_{2}|a_2\rangle)(\gamma_{1}|R_2\rangle+\gamma_{2}|L_2\rangle)(\delta_{1}|b_1\rangle+\delta_{2}|b_2\rangle)|-\rangle\\&
+\alpha_{2}\beta_{1}|L_1\rangle|a_1\rangle(\gamma_{1}|R_2\rangle+\gamma_{2}|L_2\rangle)(\delta_{1}|b_1\rangle+\delta_{2}|b_2\rangle)|-\rangle\\&
-\alpha_{2}\beta_{2}|L_1\rangle|a_2\rangle(\delta_{1}|R_2\rangle+\delta_{2}|L_2\rangle)(\gamma_{1}|b_1\rangle+\gamma_{2}|b_2\rangle)|+\rangle.
\end{split}
\end{eqnarray}

 We lastly measure the output state of the diamond NV center in the basis $\{(|+\rangle\pm|-\rangle)/\sqrt{2}\}$. If the diamond NV center in the state $(|+\rangle+|-\rangle)/\sqrt{2}$, a single-qubit operation $\sigma_z$ is performed on the photon emitted from the spatial mode $a_2$. If the diamond NV center in the state $(|+\rangle-|-\rangle)/\sqrt{2}$, no feed-forward operation is applied. After the above operations, the state of the composite system is projected onto
\begin{eqnarray}       \label{eq18}
\begin{split}
|\varphi_6\rangle=\;&
\big[\alpha_{1}\beta_{1}|R_1\rangle|a_1\rangle+\alpha_{1}\beta_{2}|R_1\rangle|a_2\rangle+\alpha_{2}\beta_{1}|L_1\rangle|a_1\rangle\big](\gamma_{1}|R_2\rangle+\gamma_{2}|L_2\rangle)(\delta_{1}|b_1\rangle+\delta_{2}|b_2\rangle)\\&
+\alpha_{2}\beta_{2}|L_1\rangle|a_2\rangle(\delta_{1}|R_2\rangle+\delta_{2}|L_2\rangle)(\gamma_{1}|b_1\rangle+\gamma_{2}|b_2\rangle).
\end{split}
\end{eqnarray}
Therefore, the quantum circuit shown in Fig. 3 completes a two-photon four-qubit Fredkin gate which swaps the polarization and spatial states of the target photon when the control photon is in the state $|L_1\rangle|a_2\rangle$, and remain in their initial states otherwise.


\section{Discussion}\label{sec3}

The most popular universal libraries for multiqubit system are $\{$Toffoli or Fredkin gate and Hadamard gates$\}$.
Implementations of the Toffoli and Fredkin gates have been attracted strong interesting in recent years, and most of them  focus on the three-qubit system with one DoF. The traditionally three-qubit Toffoli and Fredkin gates have been proposed in various systems, such as hybrid photon-matter\cite{Hyper-Toffoli-Wang,Toffoli-hybrid}, linear optics\cite{Toffoli-linear2}, single-photon\cite{ourphotongate-NV,Toffoli-photon1}, nuclear magnetic resonance\cite{Toffoli-NMR}, ion trap\cite{Toffoli-ion}, atom\cite{Toffoli-atom}, quantum dot\cite{Toffoli-QD1}, diamond NV center\cite{ourgates-NV} and superconducting\cite{Toffoli-super1,Toffoli-super2}.
In this work, we have proposed two schemes for implementing two-photon four-qubit Toffoli and Fredkin gates via NV-cavity interactions. The individual interactions between computational qubits are achieved by employing linear optical elements and diamond NV centers. The diamond NV center can be created with excellent optical coherence property, i.e., the zero-phonon line (ZPL) exhibiting the spectral diffusion close to the optical line width ($\sim$MHZ)\cite{excellent-optical}. The enhancement of the ZPL emission for the diamond NV center is particularly important as the diamond NV center can be served as a photon-emitter for quantum information process, and it has been enhanced from 3-4\%  to 70\% of the total emission\cite{enhancement-ZPL}.

In summary, we have designed two compact quantum circuits for determinately implementing two-photon four-qubit Toffoli and Fredkin gates through cavity-assisted interactions. Compared to the traditional Toffoli and Fredkin gates with one DoF, the gates presented are acting on both the polarization DoF and the spatial DoF, are robust against the decoherence caused by their environments, and reduce the photons from four to two. The traditional two-body interactions in one DoF do not appear in naturally physical systems and can not be implemented by solely using linear optical elements.  The two-body interactions involved in one single photon, which necessary for our Toffoli and Fredkin gates, can be achieved by employing linear optics. Additional photons, necessary for the cross-Kerr-based, cluster-based, or parity-check based quantum computing, are not required in our schemes, and the diamond NV center only provides a platform for photon-photon interactions. Compared to the ones synthesized by CNOT and single-qubit gates, our schemes are not only robust against the environment noises but also reduce the complexity in experiment. The length of the synthesis of a four-qubit Toffoli gate is 16 nontrivial two-qubit gates \cite{Universal}. The average fidelities and efficiencies show that our schemes may be possible with current technology.


\section{Methods}\label{sec4}

{\bf The average fidelities and efficiencies of the Toffoli and Fredkin gates.}
The key element of our schemes is the NV-cavity emitter. In the idealistic case, the optical transition rules of the emitter is perfect, however, the phase and amplitude of the incident photon are  not perfect in experiment, and they are the functions of the $\kappa$, $\kappa_s$, and $g$. If the side leakage is taken into account, there may be the reflection (transmission) part with coefficient $r_0$ ($t$) of the incident photon in the cold (hot) cavity. Therefore, in the realistic case, the optical transition rules of the incident photon described by Eq. (\ref{eq3}) should be changed with
$|R,a\rangle|+1\rangle \rightarrow r|L,d\rangle|+1\rangle + t|R,b\rangle|+1\rangle$,
$|R,a\rangle|-1\rangle \rightarrow t_0|R,b\rangle|-1\rangle + r_0|L,d\rangle|-1\rangle$,
$|L,c\rangle|+1\rangle \rightarrow r|R,b\rangle|+1\rangle + t|L,d\rangle|+1\rangle$,
$|L,c\rangle|-1\rangle \rightarrow t_0|L,d\rangle|-1\rangle + r_0|R,b\rangle|-1\rangle$,
$|R,c\rangle|+1\rangle \rightarrow t_0|R,d\rangle|+1\rangle + r_0 |L,b\rangle|+1\rangle$,
$|R,c\rangle|-1\rangle \rightarrow r|L,b\rangle|-1\rangle + t|R,d\rangle|-1\rangle$,
$|L,a\rangle|+1\rangle \rightarrow t_0|L,b\rangle|+1\rangle + r_0|R,d\rangle|+1\rangle$, and
$|L,a\rangle|-1\rangle \rightarrow r|R,d\rangle|-1\rangle + t|L,d\rangle|-1\rangle$.
 It is note that the above undesirable parts will degrade the performance of the emitter. Hence, in order to quality the performances of our schemes, we should investigate the average fidelity, which is defined as $\overline{F}=|\overline{\langle \psi_r | \psi_i\rangle}|^2$ in experiment. Here the overline indicates average over all possible input (output) states.  $|\psi_r\rangle$ and $|\psi_i\rangle$ are the normalized output states of the system in the realistic and idealistic cases, respectively. Therefore, the average fidelity of our Toffoli or Fredkin gate, $\overline{F}_T$ or $\overline{F}_F$, can be calculated by
\begin{eqnarray}       \label{eq19}
\begin{split}
\overline{F}=&\frac{1}{(2\pi)^4}\int_{0}^{2\pi}d\alpha\int_{0}^{2\pi}d\beta\int_{0}^{2\pi}d\gamma\int_{0}^{2\pi}d\delta \;   |\langle\psi_{r}|\psi_{i} \rangle|^2,
\end{split}
\end{eqnarray}
where $\cos\alpha=\alpha_1$,  $\sin\alpha=\alpha_2$,
$\cos\beta=\beta_1$,  $\sin\beta=\beta_2$,
$\cos\gamma=\gamma_1$,  $\sin\gamma=\gamma_2$, and
$\cos\delta=\delta_1$,  $\sin\delta=\delta_2$ are taken for Eqs. (\ref{eq4}-\ref{eq18}). The average fidelity of our Toffoli and Fredkin gates as functions of $g^2/\kappa\gamma$ and $\kappa_s/\kappa$ are plotted in Fig. 4, respectively.

\begin{figure}
\centering
\includegraphics[width=6 cm,angle=0]{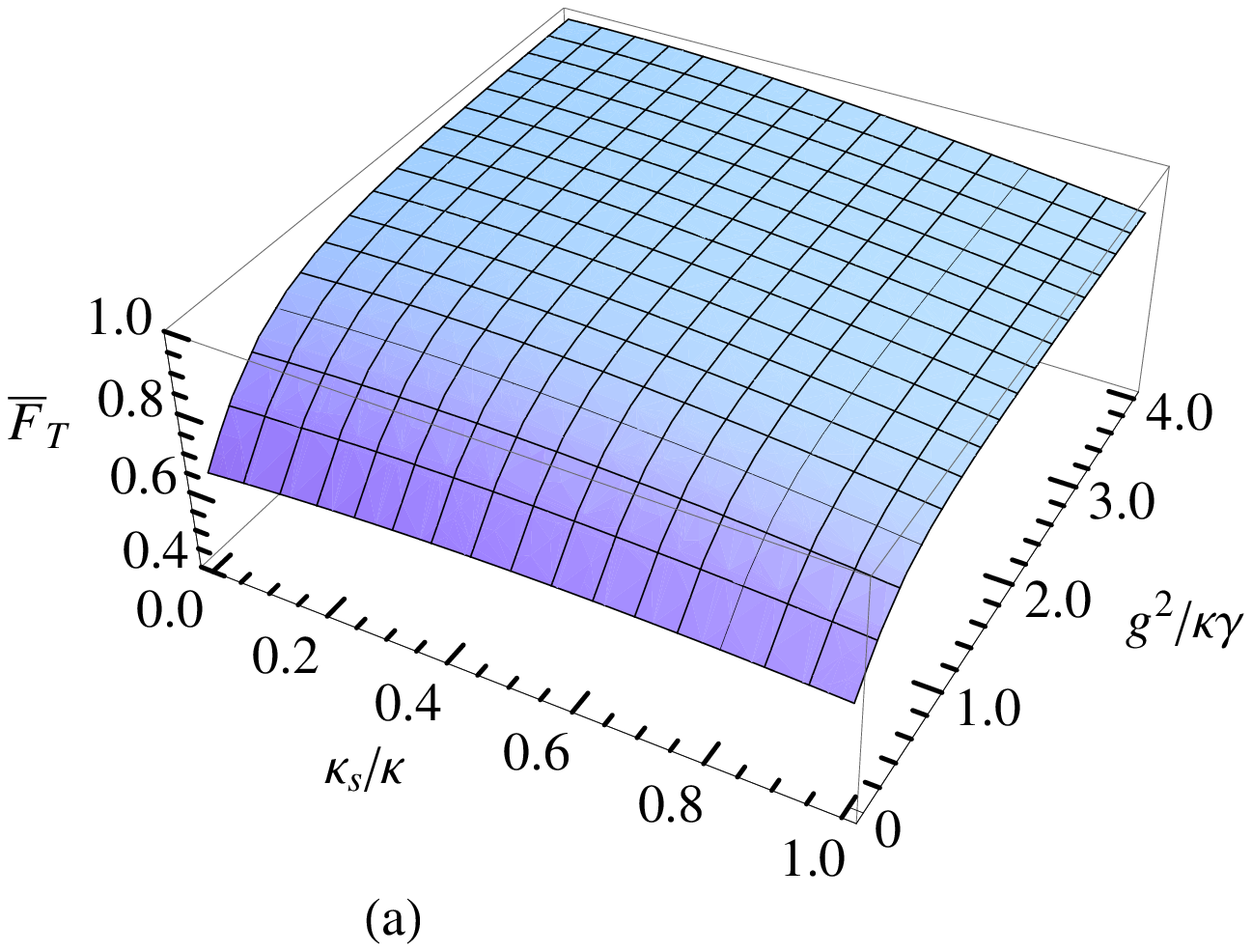}
\;\;\;\;
\includegraphics[width=6 cm,angle=0]{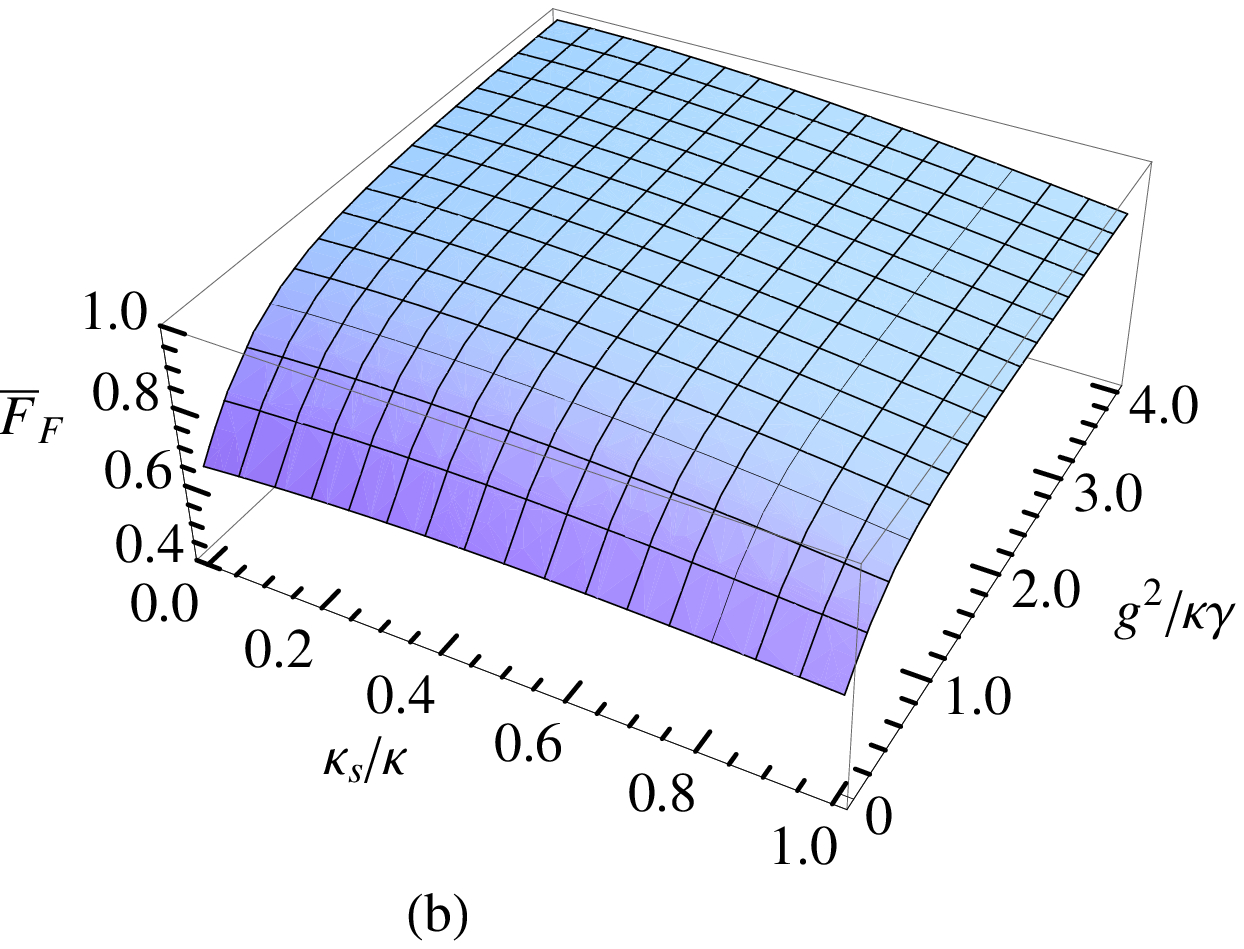}
\caption{The average fidelities averaged over all the input (output) states for (a), Toffoli gate $\overline{F}_{T}$ and (b), Fredkin gate $\overline{F}_{F}$.
} \label{Fig4}
\end{figure}

Efficiency is another powerful and practical tool for qualifying the performance of the setup. In experiment, the efficiency is defined as $n_{\rm{output}}/n_{\rm{input}}$, i.e., the yield of the incident photons. $n_{\rm{output}}$ and $n_{\rm{intput}}$ are the numbers of gate's output and input photons, respectively. The average efficiencies of our Toffoli and Fredkin gates, $\overline{\eta}_T$ and $\overline{\eta}_F$, are calculated by
\begin{eqnarray}       \label{eq20}
\begin{split}
\overline{\eta}=&\frac{1}{(2\pi)^4}\int_{0}^{2\pi}d\alpha\int_{0}^{2\pi}d\beta\int_{0}^{2\pi}d\gamma\int_{0}^{2\pi}d\delta \;   \frac{n_{\rm{output}}}{n_{\rm{intput}}},
\end{split}
\end{eqnarray}
where $\cos\alpha=\alpha_1$, $\sin\alpha=\alpha_2$,
$\cos\beta=\beta_1$,  $\sin\beta=\beta_2$,
$\cos\gamma=\gamma_1$,  $\sin\gamma=\gamma_2$, and
$\cos\delta=\delta_1$,  $\sin\delta=\delta_2$. By calculating,  one can find that
\begin{eqnarray}       \label{eq20}
\begin{split}
&\overline{\eta}_T=\overline{\eta}_F=\frac{9+3(|r_0|-|t_0|)^2+(3+(|r|-|t|)^2)(1-|t|-|r_0|)^2}{16}.
\end{split}
\end{eqnarray}
The date shows $\overline{\eta}_T$ ($\overline{\eta}_F$) against $g^2/\kappa\gamma$ and $\kappa_s/\kappa$ are presented in Fig. 5.

\begin{figure}          
\centering
\includegraphics[width=6 cm,angle=0]{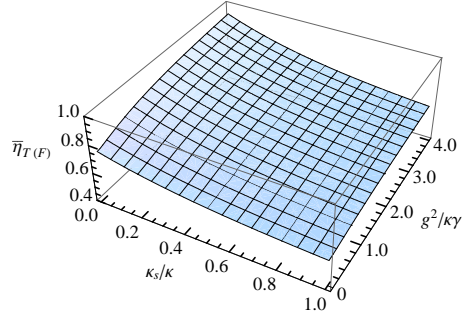}
\caption{The average efficiencies averaged over all the input (output) states for Toffoli gate $\overline{\eta}_{T}$ and Fredkin gate $\overline{\eta}_{F}$.
} \label{Fig5}
\end{figure}

 From Figure 4-5, one can see that the normalized side leakage and the normalized cavity coupling strength impact the gate's fidelity and efficiency. The higher $g^2/\kappa\gamma$ and the lower $\kappa_s/\kappa$, the higher fidelity and efficiency. The high fidelities and efficiencies of our Toffoli and Fredkin gates can be achieved. For example, in the case $\kappa_s/\kappa=0.1$ and $g^2/(\kappa\gamma)=2.4$, $\overline{F}_T=0.980436$, $\overline{F}_F=$0.979516 with $\overline{\eta}=0.847014$, and
 $\kappa_s/\kappa=1$ and $g^2/(\kappa\gamma)=2.4$, $\overline{F}_T=0.884273$, $\overline{F}_F=$0.868208 with $\overline{\eta}=0.63922$.

Our schemes are deterministic if the unavoidable effect of photon loss and  imperfections are neglected.  Our schemes in practice working with photon loss, non-ideal single-photon sources, unbalanced CPBSs and BSs, timing jitter, spin flips during the optical excitation,  the mix between state $|A_2\rangle$  and other excited states, signal to noise in the ZPL channel, imperfect spin initialization and detection. These imperfections degrade the fidelities and efficiencies of our two gates, and the imperfections which are due to the technical imperfection will be improved with further technical advances. Given the current technology, our schemes are more efficient than the traditional ones with one DoF,  the synthesis-based ones which are rely on CNOT gates and single rotations, the parity-check-based ones, and they may be possible in experiment.

\section*{Acknowledgments}

This work is supported by the National Natural Science Foundation of
China under Grant Nos. 11604012 and 11547138, and the Fundamental Research Funds for
the Central Universities under Grant No. 06500024.

\end{document}